\pdfoutput=1
\documentclass[]{article}

\usepackage{cite}
\usepackage{amssymb,amsmath,amsthm}
\usepackage[a4paper, scale = 0.8]{geometry}
\usepackage[colorlinks,linkcolor=blue,anchorcolor=blue,citecolor=red]{hyperref}
\usepackage[dvipdfm]{graphicx}
\usepackage{float}
\usepackage{subfigure}
\usepackage{tikz}

\newtheorem{myTheorem}{Theorem}
\newtheorem{myAssumption}{Assumption}
\newtheorem{myDefinition}{Definition}
\newtheorem{myLemma}[myTheorem]{Lemma}
\newtheorem{myRemark}[myTheorem]{Remark}
\newtheorem{myCorollary}[myTheorem]{Corollary}
\newtheorem{myExample}[myTheorem]{Example}

\newtheorem{myProposition}[myTheorem]{Proposition}
\newenvironment{myGoodTheorem}[2]{\begin{myTheorem}[#1]\label{theorem:#2}}{\end{myTheorem}}

\newenvironment{myGoodDefinition}[2]{\begin{myDefinition}[#1]\label{definition:#2}}{\end{myDefinition}}
\newenvironment{myGoodExample}[2]{\begin{myExample}[#1]\label{example:#2}}{\end{myExample}}
\newenvironment{myGoodProposition}[2]{\begin{myProposition}[#1]\label{proposition:#2}}{\end{myProposition}}
\newenvironment{myGoodCorollary}[2]{\begin{myCorollary}[#1]\label{corollary:#2}}{\end{myCorollary}}
\newenvironment{myGoodLemma}[2]{\begin{myLemma}[#1]\label{lemma:#2}}{\end{myLemma}}
\newenvironment{myGoodRemark}[2]{\begin{myRemark}[#1]\label{remark:#2}}{\end{myRemark}}
\newenvironment{myGoodProof*}{\begin{proof}}{\end{proof}}

\newif\ifthisisdraft
\newcommand{\mynewsentence}{\ifthisisdraft {\newline *} \fi}

\newenvironment{myblock}[1]{\ifthisisdraft \paragraph{{\color{red}>>>>#1<<<<}}-\mynewsentence \fi}{}

\newcommand{\Ito}{It\^{o}}

\newcommand{\Doleans}{Dol\'{e}ans}

\newcommand{\RadonNikodym}{Radon-Nikod\'{y}m}

\DeclareMathOperator{\myd}{d\!}

\DeclareMathOperator{\CVaR}{CV@R}

\DeclareMathOperator{\sign}{sign}
\DeclareMathOperator{\closed}{cl}

\DeclareMathOperator*{\essinf}{ess~inf}

\DeclareMathOperator*{\esssup}{ess~sup}

\newcommand{\II}{\mathbb{I}}

\newcommand{\myangle}[1]{\left\langle {#1}\right\rangle}

\newcommand{\mysint}{\!\circ\!}

\newcommand{\myVert}[1]{\left\lVert {#1}\right\rVert}
\newcommand{\myvert}[1]{\left\lvert {#1}\right\rvert}

\newcommand{\rd}[1]{\mathcal {D} ( {#1})}

\newcommand{\rc}[1]{\mathcal {C} ( {#1})}

\newcommand{\Eg}{\mathcal{E}^g}

\newcommand{\Ct}[2]{\mathcal {C} _{#1}\!\left( {#2} \right)}
\newcommand{\Dt}[2]{\mathcal {D} _{#1}\!\left( {#2} \right)}
\newcommand{\Egt}[2]{\mathcal {E} ^g_{#1}\!\left( {#2} \right)}
\newcommand{\EEt}[2]{\mathbb {E} \!\left( {#2}\Big | \mathcal {F}_{#1}\right)}

\newcommand{\Vt}[2]{\mathcal {V} _{#1}\!\left( {#2} \right)}

\everymath{\displaystyle}

\thisisdraftfalse
\title{On the Correspondence and the Risk Contribution for Conditional Coherent and Deviation Risk Measures}

\author{
        Guangyan JIA\footnote{Zhongtai Securities Institute for Financial Studies, Shandong University, Jinan, China. Email: jiagy@sdu.edu.cn},  
        Mengjin ZHAO\footnote{Zhongtai Securities Institute for Financial Studies, Shandong University, Jinan, China. Corresponding author. Email: zhaomj@mail.sdu.edu.cn}}
\date{\today}

\begin{document}
\maketitle
\begin{abstract}
	We give an axiomatic framework for conditional generalized deviation measures.
	Under financially reasonable assumptions, we give the correspondence between conditional coherent risk measures and generalized deviation measures.
	Moreover, we establish the notion of continuous-time risk contribution for conditional coherent risk measures and generalized deviation measures.
	With the help of the correspondence between these two different types of risk measures, we give a microscopic interpretation of their risk contributions.
	Particularly, we show that the risk contributions of time-consistent risk measures are still time-consistent.
	We also demonstrate that the second element of the BSDE solution $(Y, Z)$ associated with $g$-expectation has the meaning of risk contribution. 
\end{abstract}
\section{Introduction}\label{Section:Introduction}
	\begin{myblock}{Deviation and Coherent Measures}
		Different types of risk measures have been proposed for financial positions in the past tens of years.
		Among these risk measures, two types are significant: coherent risk measures and generalized deviation measures.
		The first one is introduced by \cite{artzner1999coherent} in an axiomatic framework and can be understood as the robust loss of a financial position, such as conditional value at risk(CV@R).
		The second is introduced in \cite{rockafellar_generalized_2006} to give the degree of deviation for random variables.
		Risk measures of this type can reflect the way random variables deviate from the mean.
		For example, standard deviation can be classified as generalized deviation measures.
		Actually, there is a correspondence between these two types of risk measures which is also proposed in \cite{rockafellar_generalized_2006}.
		Namely, a coherent risk measure can introduce a generalized deviation measure via this correspondence and vice versa.
	\end{myblock}

	\begin{myblock}{Risk Contribution}
		The term "risk contribution" is a tool for monitoring the risk of individual assets in a portfolio.
		The $i$-th asset's risk contribution is defined as the risk increment brought by the capital allocated to this $i$-th asset.
		Different assets will show uneven risk contributions, or, in other words, risk concentration.
		This concept comes from \cite{litterman1996hot} where the assets with higher risk increments are called ``hot spots''.
		For positive homogeneous risk measures, this tool can also provide us with a means of decomposing the total risk when measuring the risk of a portfolio.
		In this manner, the total risk can be distributed among the assets in this portfolio.
		Briefly speaking, the sum of risk contributions for component assets is the total risk of the portfolio.
		On the other hand, the risk contribution depends on its associated risk measure.
		For coherent risk measures, their related risk contributions are the increments of robust loss.
		Under generalized deviation measures, the risk contributions of assets are the deviations brought by the component assets.
	\end{myblock}

	\begin{myblock}{Time-Consistency}
		When we construct portfolios over a continuous horizon $[0,T]$, things about risk measures and risk contributions seem to be troublesome.
		Investors have different judgments about the risk of a financial position at different times.
		In the case where the information at time $t\in [0,T]$ is described by a $\sigma$-algebraic $\mathcal [F]_t$, conditional risk measures are better for monitoring the risk at the time point $t$.
		Moreover, a key problem in decision making is time-consistency --- "How can I be sure that tomorrow's decisions will satisfy today's objectives?"
		As a special case of conditional risk measures, time-consistent risk measures have a backward deduction property, and time-consistent risk measures are also known as dynamic risk measures.
		In this continuous-time setting, the correspondence and risk contribution should be extended for conditional risk measures.
	\end{myblock}

	\begin{myblock}{Literature}
		Risk-based investments can date back to the Nobel-prize-winning work of Markowitz \cite{markowitz1952portfolio} where the risk is measured by the variance of return rates.
		After that, different families of risk measures have been proposed in the literature.
		Coherent risk measures were initiated by \cite{artzner1999coherent} with their axiomatic tools and is used to quantify the robust loss of financial positions.
		\cite{detlefsen2005conditional} proposed conditional and dynamic convex risk measures of which coherent risk measures are special cases with positive homogeneity.
		Indexing the ``nonconstancy'' of random variables as their uncertainty, generalized deviation risk measures were reached in \cite{rockafellar_generalized_2006} and the correspondence to static coherent risk measures was also introduced.
		\cite{pistorius2017dynamic} gave the dynamic extension for generalized deviation measures based on the expression of conditional variance and axioms in \cite{rockafellar_generalized_2006}.
		For the recursiveness, time-consistent risk measures are connected to a nonlinear expectation, $g$-expectation introduced by \cite{peng1997backward}, which are derived from backward stochastic differential equations with the loss of financial positions as the terminal data.
		We refer to \cite{coquet2002filtration,gianin2006risk} for details.

		The term risk contribution is conceptual.
		\cite{litterman1996hot} discribed how to identify the primary sources of risk in portfolios based on the marginal impacts on total risk from small percentages of increases in each position.
		As far as we know, \cite{tasche1999risk} is the first paper to give the formal definition of risk contribution in an analytical way.
		For coherent risk measures, risk contribution is often connected to the term capital allocation for the meaning of robust loss \cite{kalkbrener2005axiomatic,cherny2009capital,cherny_two_2011} in the static or discrete-time case.
		For the applications on risk contribution which is also known as risk budgeting, we refer to \cite{qian2005risk,qian2005financial,maillard2010properties,bellini2021risk}.
		\cite{zhao2020continuous} studied the continuous-time risk contribution for terminal variance, which can be considered a special case for static generalized deviation measures and they obtained the volatility-managed portfolios introduced in \cite{moreira2017volatility} through a risk budgeting optimization.	
	\end{myblock}
	
	\begin{myblock}{Main Contribution(待定)}
		The aim of this paper is to analyze in detail the correspondence and risk contributions for coherent risk measures and generalized deviation measures in the continuous-time case.
		\begin{itemize}
				\item Now that there is the correspondence between static coherent risk measures and generalized deviation measures, we give the axiomatic description for conditional generalized deviation measures and the correspondence to conditional coherent risk measures under financially reasonable assumptions.
				As a special case, the time-consistency of generalized deviation measures is characterized.
				\item Risk contributions for these two types of conditional risk measures are derived in an analytical way.
				It is important to note that the risk contribution is continuously distributed even under a static risk measure.
				Particularly, the risk contributions for time-consistent risk measures are proved to be still time-consistent.
				\item Moreover, we also interpret the second component of the solution pair $(Y,Z)$ to BSDEs as the risk contribution of the coherent risk measures and generalized deviation measure associated with $g$-expectations.
		\end{itemize} 
	\end{myblock}

	\begin{myblock}{Structure(待定)}
		The structure of this paper is organized as follows.
		In Section~\ref{Section:Risk Measures}, we review the static risk measures and their correspondence.
		Then we give the axiomatic description for conditional generalized deviation measures.
		In Section~\ref{Section:Correspondence}, we characterize the time-consistency of conditional generalized deviation measures according to the time-consistency of coherent risk measures.
		Through constructing volatility recorders, we then give a representation of time-consist generalized deviation risk measures by $g$-expectation.
		In Section~\ref{Section:RiskContribution}, the risk contributions for conditional risk measures are derived from the subdifferential of these measures.
		We also show that the risk contributions admit an aggregation property.
		In Section~\ref{Section: Time-Consistent Risk Contribution}, the risk contributions for time-consistent risk measures are still proved to be time-consistent.
		For the recursive nature, we connect time-consistent risk contributions to the second component of the solution pair $(Y,Z)$ to BSDEs.
		$Z$ is linked to the sum of the instataneous risk contribution of assets.
		Finnally in Section~\ref{Section: Conclusion}, we give a conclusion through two figures illustrating the relations of risk-based functionals appearing in this paper and the risk contributions.  
	\end{myblock}

\section{Coherent Risk Measures and Generalized Deviation Measures}\label{Section:Risk Measures}
	\begin{myblock}{The Paragraph after Section: Basic Settings of the Static Case}
		The sample space $\Omega$ can be considered as a fixed set of scenarios.
		The random variables here serve as the financial positions in this case, i.e., the mappings $X: \Omega \to \mathbb {R}$ are the net worth after the trading under different scenarios $\omega$.
		The aim of risk measurement is to give a number $\rho (X)$ to quantify the risk of the position $X$.
		Let $(\Omega, \mathcal {F}, \mathbb {P})$ be the random basis where $\mathbb {P}$ is the reference probability measure.
		We work with the space $L^2 = L^2(\Omega,\mathcal {F}; \mathbb {P})$ in which the random variables are equipped with the norm $\myVert{X}_{L^2} = \left(\mathbb {E}X^2\right)^{1/2}$.   
	\end{myblock}
	
\subsection{Lookback on the Static Case}
	\begin{myblock}{解释Coherent和 deviation}
		There are two basic types of risk measures, coherent risk measures and generalized deviation measures.
		The first one can describe the loss of financial positions.
		CV@R, also known as the expected shortfall, is bracketed in this type.
		The latter describes the ``non-constancy'' of random variables and the standard deviation can be categorized as generalized deviation measures in an axiomatic framework.
	\end{myblock}

	\begin{myblock}{Axioms for Static Deviation/Coherent Measures}
		\begin{myGoodDefinition}
		{Coherent and Deviation Measures, \cite{artzner1999coherent,rockafellar_generalized_2006}}
		{static-measure-axiom}
			Consider a functional $\rho: L^2 \to \mathbb {R}$.
			There are statements to characterize $\rho$ acting as a risk measure:
			\begin{description}
				\item [M1] $\rho ({X+Y})\leq \rho (X) + \rho (Y)$ for all $X$ and $Y$;
				\item [M2] $\rho (0) = 0$, and $\rho ({\lambda X}) = \lambda \rho (X)$ for all $X$ and $\lambda >0$; 
				\item [C1] $\rho (X)\geq \rho (Y)$ if $X\leq Y$;
				\item [C2] $\rho (X+C) = \rho (X) - C$;
				\item [D1] $\rho ({X+C}) = \rho (X)$ for all random variables $X$ and constants $C$;  
				\item [D2] $\rho (X) \geq 0$ for all $X$, with $\rho (X)>0$ for nonconstant random variables $X$.
				\item [D2'] $\rho (X) \geq 0$ for all $X$.
			\end{description}
			If $\rho$ satisfies (M1,M2, C1,C2), $\rho$ is said to be a coherent risk measure and denoted as $\mathcal {C}$.
			If $\rho$ satisfies (M1,M2, D1,D2'), $\rho$ is said to be a generalized deviation measure and denoted as $\mathcal {D}$.
		\end{myGoodDefinition}		
	\end{myblock}	

	\begin{myblock}{解释risk-free与weak assumption}
		In fact, generalized deviation measures defined by \cite{rockafellar_generalized_2006} meet the axioms (M1,M2, D1,D2) instead of (D2').
		We replace (D2) by (D2') here in this paper for the existence of risk-free financial positions.
		Suppose that $\xi$ is a $\mathcal {D}$-risk-free position, namely $\mathcal {D}(\xi)= 0$.
		The axiom (D2) implies that $xi$ cannot be anything but constants.
		However there are assets recognized as risk-free positions without the property (D2), for example, zero-interest coupons with stochastic interest.
		For this reason we adopt the weaker statement (D2').
		In the following part of this paper, the item generalized deviation measures are connected to the statements (M1,M2, D1,D2').
	\end{myblock}

\subsection{Conditional Risk Measures}
	\begin{myblock}{Why Continuous-time Extensions}
		One can use the risk measures to solve problems like pricing, hedging, and optimization.
		However, static risk measures do not take into account the timing of payments in a stream.
		And as time varies, both the wealth and the policy in given probabilistic models become stochastic processes.
		How to quantify the risk at different times is quite interesting.
		It is important to consider the dynamic extensions, which include the time-varying information structures.
	\end{myblock}

	\begin{myblock}{Settings on the Continuous-time Extension - Triplet}
		In the preceding part of this paper, the random basis $(\Omega, \mathcal {F}, \mathbb {P})$ will be equipped with the filtration $\mathbb {F} = (\mathcal {F}_t)_{t\in [0,T]}$ where the filtration is identified as the natural filtration generated by the Brownian motion $B$ and completed by the collection of $\mathbb P$-null sets $\mathcal N$, i.e.,
		$$
			\mathcal F^0_t = \sigma (B_s,s\leq t); \mathcal F_t = \mathcal F_t^0\vee \mathcal N
		$$
		with $\mathcal F = \mathcal F_T$.	  
		The processes we discussed in this paper are assumed to be $\mathbb F$-adapted.  
		The predictable $\sigma$-algebra for processes is denoted as $\Sigma_p$ and the stochastic integral of a process $H$ with respect to a process $X$ is denoted as $H\mysint X$.
		We use the notation $L^p = L^p(\Omega,\mathcal F;\mathbb{ P })$ for the space of random variables and $L^p_t := L^p(\Omega,\mathcal F_t;\mathbb{ P })$ for short.
	\end{myblock}
	
	\begin{myblock}{Settings on the Continuous-time Extension -  Positions and Assets}
		Given a time-varying structure, financial positions at time $t$ are assumed to be the elements in $L^2(\Omega,\mathcal {F}_t;\mathbb {P})$( $L_t^p$ for short).
		The value processes of assets are characterized as a sequence of continuous special semi-martingales in
		\[\mathcal H^2_{\mathcal S} :=
		\left \{X\left | \Big \lVert \left\langle M\right\rangle^{1/2}_T\Big \rVert _{L^2} + \Big \lVert \int_0^T |\myd A_s|\Big \rVert _{L^2}< \infty , \text{ with the carnonical decomposition }X = X_0 + A + M.\right .\right \}\]
		to make sure that values of assets are in $L^2_t$ space for each $t$.
		Without the loss of generality, the dynamics of assets are assumed to be in the following form
		\[
		\label{fml:dynamic_of_assets}
		\left\{
		\begin{aligned}
		\myd & S^{(i)}_t = b^{(i)}_t\myd t + {\sigma_t^{(i)}} \myd B_t\\
		& S^{(i)}_0 = s_0^{(i)}
		\end{aligned}
		\right., \mbox{ for } i = 1,\dots,d
		\]
		where the instantaneous diffusion is a multi-dim process $\sigma^{(i)}_t$, the $i$-th row of $\sigma _t = [\sigma^{(i,j)}_t]_{i,j = 1}^{d,m}$.
		We shall take the $\mathbb R ^d$-valued process $S = [S^{(1)}, S^{(2)},\dots, S^{(d)} ]^\top$ as the collection for convenience.
	\end{myblock}

	\begin{myblock}{Settings on the Continuous-time Extension - Policy/Control}
		The policy $u$ is assumed to be a $\mathbb R^d$-valued predictable process with $u\in \mathbb S^\infty$ where
		\[\mathbb S^\infty = \Big\{X\Big|\big \lVert X\big \rVert _{\mathbb S^\infty}:=\big \lVert\sup_{s\leq T}|X_s|\big \rVert _{L^\infty}<\infty\Big\}.\] 
		For every $t$, the random variable $u_t$ is the shares we buy at time $t$.
		The space $\mathbb S^\infty$ implies that we cannot afford to hold infinitely many shares in any case.	
	\end{myblock}
	
	\begin{myblock}{Settings on the Continuous-time Extension - Investment Process}
		The {\Ito}-type stochastic integral of the policy $u$ with respect to the assets $S$ is defined as the value process of investment $X^u$ where
		\begin{equation}\label{eq:investment}
			X^u_t =x_0+  (u ^ \top \mysint S)_t =x_0 + \int_{0}^{t}u_\tau ^\top\myd S_\tau\text{ , for } t \in [0,T]
		\end{equation}
		is the value of investment $X^u$ at time $t$ with its initial wealth $x_0$.
		With $u\in \mathbb S^\infty$ and $S \in \mathcal H ^2_{\mathcal S}$, it's easy to check $X^u$ is in $\mathcal H ^2_{\mathcal S}$ by Emery's inequality.
		Hence the financial positions $X^u_t$ generated by the policy $u$ are still in $L^2_t$ for each $t$.
	\end{myblock}

	\begin{myblock}{Why Conditional Measures}
		Considering the time-varying information structure, i.e., the filtration, investors are facing the problem: how to measure the risk using the information available at the time?
		The conditional framework for convex risk measures was developed by \cite{detlefsen2005conditional}.
		Given a sub-$\sigma$-algebra $\mathcal {G}$ of $\mathcal {F}_T$, a conditional risk measure is a mapping from $L_T^2$ to $L^2(\Omega, \mathcal {G};\mathbb {P})$.
		When we take $\mathcal {F}_t$ as the sub-$\sigma$-algebra, it is said to be a risk measure conditional at $t$.
		Conditional coherent risk measures are defined in this manner and compatible with static situations when we take $t = 0$.
	\end{myblock}	

	\begin{myblock}{Conditional Coherent Risk Measure}
		\begin{myGoodDefinition}
		{Conditional Coherent Measures, \cite{detlefsen2005conditional}}
		{conditional-coherent-measures}
			With the fixed time $t$, a mapping $\mathcal {C}_t:L_T^2 \to L^2_t$ is said to be a coherent risk measure conditional at $t$ if it satisfies the following properties.
			\begin{description}
				\item [M1t] $\mathcal {C}_t ({X+Y})\leq \mathcal {C}_t (X) + \mathcal {C}_t (Y)$ for all $X$ and $Y$ in $L_T^2$;
				\item [M2t] $\mathcal {C} (0) = 0$, and $\mathcal {C} ({\lambda X}) = \lambda \mathcal {C} (X)$ for all $X \in L_T^2$ and $0<\lambda \in L^2 _t$;
				\item [C1t] $\mathcal {C}_t (X)\geq \mathcal {C}_t (Y)$ if $X\leq Y$;
				\item [C2t] $\mathcal {C}_t (X+C) = \mathcal {C}_t (X) - C$ for all $C\in L^2 _t$.
			\end{description}
		\end{myGoodDefinition}
	\end{myblock}

	\begin{myblock}{Conditional Deviation Measure}
		For the purpose of establishing the correspondence between these two types of risk measures, here we give the definition of the conditional generalized deviation measures heuristically.
		\begin{myGoodDefinition}
		{Conditional Deviation Measures}
		{conditional-deviation-measures}
			With the fixed time $t$, a mapping $\mathcal {D}_t:L_T^2 \to L^2_t$ is said to be a generalized deviation risk measure conditional at $t$ if it satisfies (M1t, M2t) in Definition~\ref{definition:conditional-coherent-measures} and the following two properties.
			\begin{description}
				\item [D1t] $\mathcal {D}_t ({X+C}) = \mathcal {D}_t (X)$ for all random variables $X\in L_T^2$ and $C \in L^2_t$;  
				\item [D2't] $\mathcal {D}_t (X) \geq 0$ for all $X\in L_T^2$.
			\end{description}
		\end{myGoodDefinition}
	\end{myblock}

	\begin{myblock}{Remark on (D2't)}
		\begin{myGoodRemark}
		{}
		{}
			Note that we use a weaker statement (D2't), so the definition of conditional generalized deviation measures here is slightly different from that proposed in \cite{pistorius2017dynamic} where the positivity is stated as
			\begin{description}
				\item [D2t] $\mathcal {D}_t (X) \geq 0$ for all $X\in L_T^2$, and $\Ct t X = 0$ if and only if $X$ is $\mathcal {F}_t$-measurable.
			\end{description}
			In other words, the risk-free property $\Ct t X = 0$ for some $X \in L_T^2$ is allowed here.
			As for the ``only if'' part, $\Dt t X = 0$ for any $X\in L_t^2$ can be derived by
			\[2 \Dt tX = \Dt{t}{2X} = \Dt{t}{X + X} = \Dt{t}{X}.\]
		\end{myGoodRemark}	
	\end{myblock}	
	
\subsection{Time-consistent Risk Measures}\label{Subsection:Time-Consistent Measures}
	\begin{myblock}{Interpretation for time-consistent measures}
		Now we consider two families of conditional risk measures, $(\mathcal {C}_t)_{t\in [0,T]}$ or $(\mathcal {D}_t)_{t\in [0,T]}$.
		The time-consistency of these conditional risk measures is the consistency of risk assessments at different time $t\in [0,T]$.
	\end{myblock}
	\begin{myblock}{The Idea of Time-consistency for Coherent Measures}
		It is the generalized losses of financial positions that coherent risk measures charaterize.
		Hence the time-consistency for coherent risk measures implies the recursiveness of losses. 
		\begin{myGoodDefinition}
		{Time-consistency for Coherent Measures}
		{time-consistent-coherent-measures}
			A family of conditional coherent risk measures $\mathcal {C} = (\mathcal {C}_t)_t$ is said to be \emph{time-consistent} if 
			\begin{description}
				\item[C3t] $\mathcal {C}_s(X) = \mathcal {C}_s(-\mathcal {C}_t (X))$ for any $X\in L^2_T$ with $0\leq s\leq t\leq T$.
			\end{description}			
		\end{myGoodDefinition}
	\end{myblock}

	\begin{myblock}{The Idea of Time-consistency for Deviation Measures}
		As for generalized deviation measures, what they describe is the generalized ``deviation amplitude'' for financial positions.
		The time-consistency should describe the backward accumulation of ``deviation amplitude'' from the terminal time.
		Based on this idea, we propose the time-consistency for generalized deviation measures.
		\begin{myGoodDefinition}
		{Time-consistency for Deviation Measures}
		{time-consistent-deviation-measures}
			A family of conditional generalized deviation risk measures $\mathcal {D} = (\mathcal {D}_t)_t$ is said to be \emph{time-consistent} if 
			\begin{description}
				\item[D3t] $\mathcal {D}_s(X) = \EEt {s}{\Dt{t}{X}}+ \Dt{s}{\EEt{t}{X} - \Dt{t}{X}}$ for any $X\in L^2_T$ with $0\leq s\leq t\leq T$.
			\end{description}			
		\end{myGoodDefinition}
		This definition can be derived from the time-consistency of the coherent risk measures for the correspondence 
		\[\Dt{t}{X} = \Ct{t}{X} - \EEt{t}{-X}\]
		in Theorem~\ref{theorem:conditional-correspondence}.

		Notice that the outcome of the measurement is decomposed into two parts in the expression
		\begin{equation}
		\label{eq:time-consistent-deviation}
			\mathcal {D}_s(X) = \EEt {s}{\Dt{t}{X}}+ \Dt{s}{\EEt{t}{X} - \Dt{t}{X}}
		\end{equation}
		where the left-hand side can be read as the deviation accumulated from $T$ to $s$.
		The first term in the right-hand side is the conditional expectation at $s$ of the deviation accumulated from $T$ to $t$ and can be recognized as $[t,T]$-part of the left-hand side.
		What has been measured in the second term on the right-hand side is the $\mathcal {F}_t$-conditional expectation of $X$ deducting the $[t,T]$-part deviation.
		This second term can be recognized as the $[s,t]$-part of the total deviation accumulated over $[s,T]$.
		In terms of deviation accumulation, it is reasonable to propose the time-consistency for generalized deviation measures in this manner.

		On the other hand, this definition is compatible with the time-consistency of the coherent risk measure in the view of correspondence in Section~\ref{Section:Correspondence}.
		The correspondence
		\[\Ct{t}{X} = - \EEt{t}{-X} + \Dt{t}{X} \]
		established in Theorem~\ref{theorem:conditional-correspondence} can also imply the correspondence of time-consistency for these two types of risk measures.
		Replacing $\Ct{t}{X}$ in Definition~\ref{definition:time-consistent-coherent-measures} by $- \EEt{t}{-X} + \Dt{t}{X}$, we get
		\begin{equation}
		\EEt{s}{X} - \Dt{s}{X} = \EEt{s}{\EEt{t}{X} - \Dt{t}{X}} - \Dt{s}{\EEt{t}{X} - \Dt{t}{X}}
		\end{equation}
		which implies the time-consistency for generalized deviation measures in Definition~\ref{definition:time-consistent-deviation-measures} and vice versa.	
	\end{myblock}

\section{Representation and Correspondence for Risk Measures}\label{Section:Correspondence}
	\begin{myblock}{Introduce the dual representation}
		Recall the work \cite{artzner1999coherent} that static coherent risk measures admit the dual representation (we take the $L^2$ version here)
		\[
			\rc X = \sup_{Q\in \mathcal {Q} } \mathbb {E}[-XQ]
		\]
		where $\mathcal {Q}$ is a collection of {\RadonNikodym} derivetives with respect to $\mathbb {P}$.
		The representation above implies that coherent risk measures are robust expectations of the loss.
		After that the representation for static generalized deviation measures is proposed by
		\[ \rd X = \sup_{Q\in \mathcal {Q}}\mathbb {E}[-XQ] - \mathbb {E}[-X]\]
		in \cite{rockafellar_generalized_2006}.
		The set of those {\RadonNikodym} derivetives is called risk envelope and the probabilities introduced by the elements in $\mathcal {Q}$ are called candidate probabilities.
		Based on the dual representations, the correspondence between these two types of static risk measures is also proposed.

		Before giving the representation for conditional deviation measures, we need to make two assumptions.
		\begin{description}
			\item [A1] Conditional generalized deviation measures are lower range dominated, i.e.,
				\[\Dt{t}{X}\leq \EEt{t}{X} - \essinf X \text{ for all } X.\]			
		\end{description}
		The assumption (A1) states that deviation measures cannot exceed the lower range of $X$, $\esssup \left [\EEt{t}{X} - X\right ]$, which can be considered as the largest downside risk from the reference level $\EEt t X$.	

		\begin{description}
			\item [A2] The reference probability $\mathbb {P}$ is also a candidate probability.
		\end{description}
		This assumption can be interpreted in two ways.
		When we get the corresponded risk envelope by a risk measure, the current reference probability should be in this set of probability measures; when we give the risk measure by a family of probabilities, we require the reference probability to be in this set of probabilities.
		For example, a coherent risk measure (static or conditional) always appears with a set of candidate probabilities.
		It is well known that $\mathcal {Q}_\lambda := \{Q \mid Q\geq 0, \mathbb {E}Q=1, Q \leq \dfrac{1}{\lambda}\}$ can give the conditional value at risk $\CVaR$ at level $\lambda \in (0,1]$ by $\CVaR (X):= \sup_{Q\in \mathcal {Q}_\lambda} \mathbb {E}[-QX]$.
		We can see that $1 \in \mathcal {Q}_\lambda$, which  satisfies this assumption.
		In this view, it seems that the assumption (A2) is quite reasonable: the reference probability $\mathbb P$, generally acting the probability of the real world, should be taken into account as a candidate when taking the robust expected loss of financial positions.		
		The assumption (A2) implies
		\[\Ct{t}{X}\geq \EEt{t}{-X} \text{ and }\Dt{t}{X}\geq 0\text{ for any }X\in L^2_T\]
		where $\mathcal {C}_t, \mathcal {D}_t$ are $\mathcal {F}_t$-conditional risk measures we study in this paper.
	\end{myblock}
\subsection{Dual Representation for Conditional Generalized Deviation Measures}
	\begin{myblock}{Dual Representation}
		\begin{myGoodTheorem}
		{Dual Representation}
		{conditional-deviation-representation}   
		Under assumptions (A1) and (A2), a functional $\mathcal {D}_t : L_T^2 \to L_t^2$ is a closed conditional generalized deviation measure if and only if it has a representation of the form
		\begin{equation}
			\mathcal {D}_t(X) = \mathbb {E}[X|\mathcal {F}_t] - \essinf _{Q\in \mathcal {Q}_t}\mathbb {E}[XQ| \mathcal {F}_t] = \esssup _{Q\in \mathcal {Q}_t}\mathbb {E}[-XQ| \mathcal {F}_t] - \mathbb {E}[-X|\mathcal {F}_t]
		\end{equation}
		where the risk envelope $\mathcal {Q}_t$ has the properties
		\begin{itemize}
			\item $\mathcal {Q}_t$ is a non-empty closed convex subset of $L_T^2$;
			\item the elements $Q$ in $\mathcal {Q}_t$ satisfy $Q\geq 0,  \mathbb {E}Q = 1$;
			\item probabilities introduced by $\myd \mathbb {Q}  = Q \myd \mathbb {P}$ with $Q\in \mathcal {Q}_t$ are consistent on $\mathcal {F}_t$, namely $\mathbb {Q}\big | _{\mathcal {F}_t} = \mathbb {P}\big | _{\mathcal {F}_t}$.
		\end{itemize}
		Here the closedness in this statement is that of the risk measure when acting as a convex operator, which is equivalent to the upper semi-continuity of it.
		\end{myGoodTheorem}

		\begin{myGoodProof*}{}{}
			[``if''-part] (M1t)(M2t) is obvious.
			Notice that $\mathbb {Q}\big | _{\mathcal {F}_t} = \mathbb {P}\big | _{\mathcal {F}_t}$ with $\myd \mathbb {Q} = Q\myd \mathbb {P}$.
			With the consistency on $\mathcal {F}_t$, we get 
			\[\esssup _{Q\in \mathcal {Q}_t}\EEt{t}{-(X+C)Q} = \esssup _{Q\in \mathcal {Q}_t}\EEt{t}{-XQ}-C\] for any $C\in L_t^2$ and $X\in L_T^2$.
			Then we have 
			\[\esssup _{Q\in \mathcal {Q}_t}\EEt{t}{-(X+C)Q} - \EEt{t}{-(X+C)} = \esssup _{Q\in \mathcal {Q}_t}\EEt{t}{-XQ} - \EEt{t}{-X}\]
			which implies (D1t).
			The assumption (A2) implies $\esssup _{Q\in \mathcal {Q}_t}\EEt{t}{-XQ} - \EEt{t}{-X}\geq 0$ and hence we get (D2't).

			[``only if''-part] Firstly we will construct an auxilary static coherent risk measure and give its envelope.
			Now suppose $\mathcal {D}_t$ is a closed $\mathcal {F}_t$-conditional generalized deviation risk measure in Definition~\ref{definition:conditional-deviation-measures}.
			We can define a map $\rho : L_T^2 \to \mathbb {R}$ by
			\[\rho (X):= \mathbb {E}\left[\Dt{t}{X} - \EEt{t}{X}\right].\]
			With $Y\leq X$, we have 
			\[
			\begin{aligned}
			\rho(Y) - \rho (X) &= \mathbb {E}\left[\Dt{t}{Y} - \Dt{t}{X}+ \EEt{t}{Y-X}\right]
			\\
			&= \mathbb {E}\left[\Dt{t}{Y} - \Dt{t}{X-Y+Y}+ \EEt{t}{Y-X}\right]
			\\
			& \geq \mathbb {E}\left[ \Dt{t}{Y} - \Dt{t}{X-Y} - \Dt{t}{Y}+ \EEt{t}{Y-X}\right]
			\\
			& = \mathbb {E}\left[\EEt{t}{X-Y} -\Dt{t}{X-Y} \right]
			\\
			& \geq \mathbb {E}\left[\essinf(X-Y)\right] \geq 0.
			\end{aligned}	 
			\]
			It is easy to check the left properties in Definition~\ref{definition:static-measure-axiom} so that $\rho$ is a static coherent risk measure.
			Thus from the dual representation for static coherent risk measures in \cite{artzner1999coherent}(Here we adopt the $L^2$ version.),  $\rho$ admits a representation
			\[\rho (X) = \sup_{Q\in \mathcal {Q}}\mathbb {E}[-XQ]\]
			where $\mathcal {Q}$ is a non-empty closed convex subset in $L_T^2$ with its elements satisfying $Q\geq 0$ and $\mathbb {E}Q = 1$.

			Next we will show that the elements in $\mathcal {Q}$ is consistent on $\mathcal {F}_t$.
			Based on the expression of $\rho$, we have
			\[\rho (\II _A) = \mathbb {E}\left[ \Dt{t}{\II _A} - \EEt{t}{\II _A}\right] = -\mathbb {P}(A) \text{ and } \rho (-\II _A) = \mathbb {P}(A)\]
			for arbitrary set $A \in \mathcal {F}_t$.
			On the other hand, by the representation of $\rho$ above, we also have 
			\[
				\rho (\II_A)  = \sup_{Q\in \mathcal {Q}}\mathbb {E}[-Q\II_A]
				, 
				\rho (-\II_A)  = \sup_{Q\in \mathcal {Q}}\mathbb {E}[Q\II_A].  
			\]
			Hence for any $\mathcal {F}_t$-measureble set $A$ we get 
			\[-\rho (\II _A) = -\sup_{Q\in \mathcal {Q}}\mathbb {E}[-Q\II_A] = \inf_{Q\in \mathcal {Q}}\mathbb {E}[Q\II_A] = \mathbb {P}(A), \rho(-\II_A) =\sup_{Q\in \mathcal {Q}}\mathbb {E}[Q\II_A] = \mathbb {P}(A)\]
			which implies $\mathbb {E}[Q\II_A] = \mathbb {P}(A)$.
			In other words, the probabilities $\mathbb {Q}$ introduced by $\myd \mathbb {Q} = Q\myd \mathbb {P}$ for each $Q\in \mathcal {Q}$ are consistent on $\mathcal {F}_t$.

			Finally we give the robust representation of $\mathcal {D}_t$.
			Notice that 
			\[
				-\II _AXQ = \begin{cases}
					-XQ \text{ on }A
					\\
					0 \text{ on }\Omega / A
				\end{cases}
			\]
			for each $A \in \mathcal {F}_t$ and we have
			\[
				\esssup_{Q\in \mathcal {Q}}\EEt{t}{-\II_AXQ} = \esssup_{Q\in \mathcal {Q}}\II _A \EEt{t}{-XQ}  = \II_A\esssup_{Q\in \mathcal {Q}}\EEt{t}{-XQ}
			\]
			where the first equality holds because $\II_A$ is $\mathcal {F}_t$-measurable and the second equality is valid because $\mathcal {Q}$ is consistent on $\mathcal {F}_t$.
			On the other hand we also have
			\[\sup_{Q\in \mathcal {Q}}\mathbb {E}[QX] = \sup_{Q\in \mathcal {Q}}\mathbb {E}\left[\EEt t {QX}\right] = \mathbb {E}\left[\esssup_{Q\in \mathcal {Q}}\EEt t {QX} \right] \text{ for arbitrary }X \in L_T^2\] 
			where the last equality holds for that $\mathcal {Q}$ is consistent on $\mathcal {F}_t$.
			For arbitrary $\mathcal {F}_t$-measurable set $A$, comparing 
			\[
				\mathbb {E}\left[\II_A\esssup _{Q\in \mathcal {Q}}\EEt{t}{-XQ} - \II_A \EEt{t}{-X}\right] = \sup_{Q\in \mathcal {Q}}\mathbb {E}\left[-XQ\II_A + X\II_A\right]
			\]
			with
			\[
				\rho (\II_A X) = \mathbb {E}\left[\Dt{t}{\II_AX} - \EEt{t}{\II_AX}\right] = \mathbb {E}\left[\II_A\left(\Dt{t}{X} - \EEt{t}{X}\right)\right],
			\]
			we can get 
			\[
				\esssup _{Q\in \mathcal {Q}}\EEt{t}{-XQ} = \Dt{t}{X} - \EEt{t}{X}	\text{ $\mathbb {P}$-a.s. on $\mathcal {F}_t$}.
			\]
			Because the set $\mathcal {Q}$ depends on $\mathcal {D}_t$ and its elements are consistent on $\mathcal {F}_t$, we can also denote it as $\mathcal {Q}_t$.
			Now we complete the ``only if''-part.
		\end{myGoodProof*}   
	\end{myblock}
	
\subsection{Correspondence}
	\begin{myblock}{Correspondence between Conditional Coherent and Generalized Deviation Measures}
		As two important types of risk measures, static coherent and deviation measures have their correspondence with each other as proposed by \cite{rockafellar_generalized_2006}.  
		Here we give a conditional version.
	\end{myblock}

	\begin{myblock}{Conditional Correspondence}
		\begin{myGoodTheorem}
		{Conditional Correspondence}
		{conditional-correspondence}
			Under assumptions (A1) and (A2), conditional coherent and deviation risk measures have a one-to-one correspondence with each other under the rules
			\begin{itemize}
				\item $\mathcal {D}_t$ introduced by $\Dt{t}{X} := \Ct{t}{X} - \EEt{t}{-X}$ is a $\mathcal {F}_t$-conditional deviation risk measure when $\mathcal {C}_t$ is a $\mathcal {F}_t$-conditional coherent risk measure;
				\item  $\mathcal {C}_t$ introduced by $\Ct{t}{X} := \Dt{t}{X} + \EEt{t}{-X}$ is a $\mathcal {F}_t$-conditional coherent risk measure when $\mathcal {D}_t$ is a $\mathcal {F}_t$-conditional deviation risk measure.
			\end{itemize}
		\end{myGoodTheorem}
		\begin{myGoodProof*}
		{}
		{}
			It is easy to check (M1t) and (M2t) for both two directions.
			We only need to treat (C1t)(C2t) or (D1t)(D2t).

			The proof of the first statement:
			\begin{itemize}
				\item (D1t) For each $C\in L_t^2$ and $X\in L_T^2$, we have
				\[\Dt{t}{X+C} = \Ct{t}{X+C} + \EEt{t}{X+C} = \Ct{t}{X} - C + \EEt{t}{X}+C = \Dt{t}{X}.\]
				\item (D2t') With assumption (A2), we can get 
				\[ \Dt{t}{X} = \Ct{t}{X} + \EEt{t}{X} \geq 0.\]
			\end{itemize}

			The proof of the second statement:
			\begin{itemize}
				\item (C1t) Let $X\geq Y$ and $X' := X-Y$.
				We can get 
				\[
				\begin{aligned}
				\Ct{t}{Y} - \Ct{t}{X} &= \Dt{t}{Y} - \EEt{t}{Y} - \Dt{t}{X}+ \EEt{t}{X}
				\\
				& = \Dt{t}{Y} - \EEt{t}{Y} - \Dt{t}{Y+X'}+\EEt{t}{Y+X'}
				\\
				&= \Dt{t}{Y } - \Dt{t}{Y+X'}+\EEt{t}{X'}
				\\
				& \geq \Dt{t}{Y} - \Dt{t}{Y} - \Dt{t}{X'} + \EEt{t}{X'}
				\\
				& \geq -\EEt{t}{X'} + \essinf X' + \EEt{t}{X'} \geq 0.
				\end{aligned}	 
				\]
				The last inequality is ensured by the assumption (A1).
				\item (C2t) With $C\in L_t^2$, we have 
				\[\Ct{t}{X+C} = \Dt{t}{X+C} - \EEt{t}{X+C} = \Dt{t}{X} - \EEt{t}{X} - C = \Ct{t}{X} - C.\]
			\end{itemize}
		\end{myGoodProof*}		
	\end{myblock}

	\begin{myblock}{Interpretation for Correspondence}
		In the view of dual representation, the coherent risk measures of a financial position $X$ is the robust expected loss under the candidate probabilities introduced by $\mathcal {Q}$.
		The difference between the robust expected loss $\Ct{t}{X}$ and referenced expected loss $\EEt{t}{-X}$ is the deviation measure $\Dt{t}{X}$.
		Figuratively speaking, the generalized deviation measure acts as the deviation amplitude of the referenced loss.
		Conversely, given a deviation measure, we can get an estimation of the loss by substracting the deviation amplitude from the referenced loss. 
		This estimation is a coherent risk measure.
	\end{myblock}  

\subsection{\texorpdfstring{$g$}{g}-expectation and time-consistency}
	\begin{myblock}{BSDE and \texorpdfstring{$g$}{g}-expectation}
		When it comes to recursiveness, time-consistent coherent risk measures are often associated with $g$-expectation, which is a filtration-consistent sublinear expectation introduced by backward stochastic differential equations (BSDE).
		Consider a functional $g: \Omega \times [0,T] \times \mathbb {R}^m \to \mathbb {R}$ with the assumptions
		\begin{itemize}
			\item $g$ is Lipschitz in $z$, i.e.,
			\[\myvert {g(t,\omega,z) - g(t,\omega,z')} \leq C \myVert{z-z'} \]
			for some constant $C>0$ and arbitrary $z,z' \in \mathbb {R}^m$;
			\item $\mathbb {E}\int_0^T |g_t(z)|^2\myd t <\infty$ for $z \in \mathbb {R}^m$.
		\end{itemize}
		By \cite{pardoux1990adapted}, under these assumptions on $g$, for every $X\in L_T^2$, the BSDE
		\begin{equation}
		\label{eq:BSDE}
			\begin{cases}
			-\myd Y_t = g_t(Y_t,Z_t) \myd t - Z_t \myd B_t
			\\
			Y_T = X
			\end{cases}
		\end{equation}
		admits a unique solution $(Y,Z)$.
		In \cite{peng1997backward}, the conditional $g$-expectation of $X$ is defined by
		\[\Egt{t}{X}:= Y_t\]
		where $Y$ is the first component of the solution pair $(Y,Z)$.
		According to \cite{gianin2006risk}, the associated $g$-expectation of $-X$, namely $\Eg (-X)$, is a time-consistent conditional coherent risk measure of the financial position $X\in L_T^2$.
		In other words, the $g$-expectation of the loss $-X$ is a time-consistent coherent risk measure of $X$.
		On the other hand, as a corollary of \cite{coquet2002filtration}, the time-consistent conditional coherent measures can also be represented by a $g$-expectation with a unique functional $g$.
	\end{myblock}

	\begin{myblock}{\texorpdfstring{$g$}{g}-deviation}
		To illustrate the accumulation property of the time-consistent generalized deviation measures, here we construct a robust volatility recorder which is also called $g$-deviation in \cite{pistorius2017dynamic}.
		Suppose $X$ is a financial position in $L_T^2$ and let $M_t:= \EEt{t}{X}$ for each $t$.
		By the predictable representation for the martingale $M$, $X$ can be expressed as 
		\begin{equation}
		\label{eq:martingale-representation}
		X = M_T = M_t + \int_t^T \sigma_s^X \myd B_s
		\end{equation}
		with the unique integrand $\sigma^X$ for each $t$.
		\begin{myGoodDefinition}
		{Robust Volatility Recorder}
		{volatility-recorder}
			Given a collection $\mathcal {K}$ of $m$-dim predictable process $K\in \mathbb {S}^\infty$ such that 
			\begin{equation}
			\label{eq:recorder-weight}
				\sup_{K\in \mathcal {K}}\myangle{K_t, x}\geq 0,  \forall x \in \mathbb {R}^m
			\end{equation}
			for each $t\in [0,T]$.
			The \emph{robust volatility recorder} $\mathcal {V}= (\Vt{t}{X})_t$ of $X$ introduced by $\mathcal {K}$ is a sequence of mappings $(\mathcal {V}_t)_{t\in [0,T]}$ defined by
			\begin{equation}
			\label{eq:volatility-recorder}
				\Vt{t}{X}:= \EEt{t}{\int_t^T \sup_{K\in \mathcal {K}}\myangle{K_s, \sigma^X_s}\myd s}\text{ for each }t.
			\end{equation}
		\end{myGoodDefinition}

		\begin{myGoodProposition}
		{Volatility Recorder and Deviation}
		{volatility-recorder-is-deviation}
			The robust volatility recorder in \ref{definition:volatility-recorder} is a family of conditional generalized deviation measures indexed by $t$.
		\end{myGoodProposition}
		\begin{myGoodProof*}{}{}
			We should check that $\mathcal {V}_t$ coincides with the definition of conditional deviation measures for each $t$.
			\begin{itemize}
				\item (M1t). Notice that the representation in equation~(\ref{eq:martingale-representation}) satisfies
			\[\sigma^{X+Y} = \sigma^X + \sigma^Y.\]
			Then (M1t) can be checked by
			\[\begin{aligned}
			\Vt t {X+Y} &=  \EEt{t}{\int_t^T \sup_{K\in \mathcal {K}}\myangle{K_s, \sigma^{X+Y}_s}\myd s} 
			\\
			& \leq  \EEt{t}{\int_t^T \sup_{K\in \mathcal {K}}\myangle{K_s, \sigma^X_s} + \sup_{K\in \mathcal {K}}\myangle{K_s, \sigma^Y_s}\myd s} 
			\\
			& = \Vt t X + \Vt t Y.
			\end{aligned}\]
			\item (M2t). For any $\lambda \in L_t^2$, notice that 
			\[\lambda X = \lambda M_t + \int_t^T \lambda \sigma_s^X \myd B_s\]
			which implies $\sigma ^{\lambda X} = \lambda \sigma^X$.
			Then with each $t\in [0,T]$ we get 
			\[\sup_{K\in \mathcal {K}}\myangle{K_t, \sigma^{\lambda X}_t} = \sup_{K\in \mathcal {K}}\myangle{K_t, \lambda \sigma^X_t} = \lambda \sup_{K\in \mathcal {K}}\myangle{K_t, \sigma^X_t}, \mathbb {P}\text{-a.s.}\]
			from which we have (M2t) valid
			\[\begin{aligned}
			\Vt t {\lambda X} &= \EEt{t}{\int_t^T \sup_{K\in \mathcal {K}}\myangle{K_s, \sigma^{\lambda X}_s}\myd s} 
			\\
			&= \EEt{t}{\int_t^T \lambda\sup_{K\in \mathcal {K}}\myangle{K_s, \sigma^X_s}\myd s} 
			\\
			&= \lambda\EEt{t}{\int_t^T \sup_{K\in \mathcal {K}}\myangle{K_s, \sigma^X_s}\myd s} = \lambda\Vt t { X}.
			\end{aligned}\]
			\item (D1t). For $C \in L_t^2$, just notice $\sigma^C_s = 0$ when $s\in [t,T]$ and we have
			\[\begin{aligned}
				\Vt t { X+C} &= \EEt{t}{\int_t^T \sup_{K\in \mathcal {K}}\myangle{K_s, \sigma^{ X+C}_s}\myd s} 
				\\
				&= \EEt{t}{\int_t^T \sup_{K\in \mathcal {K}}\myangle{K_s, \sigma^X_s}\myd s} 
				\\
				& =\Vt t { X}.
				\end{aligned}\]
			\item (D2't). (D2't) holds true for the property of $\mathcal {K}$ in the statement.
			\end{itemize}		
		\end{myGoodProof*}
	\end{myblock}

	\begin{myblock}{Time-consistent Deviation Measures are Vol.-recorders}
		\begin{myGoodTheorem}
		{Recorder Weights}
		{deviation-is-volatility-recorder}
			Suppose $\mathcal {D}$ is a time-consistent generalized deviation measure.
			Then $\mathcal {D}$ can be expressed as a robust volatility recorder with a unique set $\mathcal {K}$.
		\end{myGoodTheorem}
		\begin{myGoodProof*}{}{}
			The generalize deviation measure can derive its associated coherent risk measure $\mathcal {C}$ by the correspondence in Theorem~\ref{theorem:conditional-correspondence}.
			Time-consistency of $\mathcal {C}$ can also be checked with the help of correspondence and the time-consistency of $\mathcal {D}$.
			Then $\mathcal {C}$ can be represented by a $g$-expectation according to \cite{coquet2002filtration}.		  
			That is to say  
			\[\Ct{t}{X} = \Egt{t}{-X}\]
			where $\Egt{t}{X}$ is identified by the first component of the solution $(Y,Z)$ to the BSDE
			\[
				\begin{cases}
					-\myd Y_t = g_t(Z_t) \myd t - Z_t \myd B_t
					\\
					Y_T =- X.
				\end{cases}	
			\]
			Here we denote $g(\omega,t,z)$ as $g_t(z)$ in brief for that $(g_t(\cdot))_t$ is a stochastic process.
			We should notice that the generator of this BSDE is sublinear in $Z$ and can be expressed as
			\[g_t(z) = \sup_{\phi\in \Phi}\myangle{\phi_t,z}\]
			by its convex conjugate with a set $\Phi$.
			The process $Y$ has the expression
			\[Y_t = \sup_{Q\in \mathcal {Q}} \EEt t {-XQ} \text{ for each } t \in [0,T]\]
			where $\mathcal {Q}$ is given by
			\[\mathcal {Q} = \{Q \mid Q = \mathcal {E}(\phi^\top \mysint B)_T, \phi \in \Phi\}\]
			with {\Doleans}-Dade stochastic exponentials denoted by $\mathcal {E}(\phi ^\top B)$.
			We recommend Proposition~3.3.2 in \cite{delong2013backward} for details.

			Notice that the conditional expectation is given on $\mathcal {F}_t$.
			We can construct a subset $\mathcal {Q}_t$ of $\mathcal {Q}$ by 
			\[\mathcal {Q}_t := \left\{\mathcal {E}(\bar \phi^\top \mysint B)_T\Big | \bar \phi_s = \begin{cases}\phi_s, s\in[0,T]
			\\
			0, s\in [0,t)
			\end{cases} \text{ with }\phi \in \Phi\right\}.\]
			The backward recursiveness of $\mathcal {C}$ implies
			\[Y_t = \esssup_{Q\in \mathcal {Q}} \EEt t {-XQ} = \esssup_{Q\in \mathcal {Q}_t}\EEt t {-XQ}\]
			and \[\mathcal {Q} \supseteq \mathcal {Q}_{t_1}\supseteq \mathcal {Q}_{t_2}\supseteq \mathcal {Q}_T=\{1\}, \forall t_1\leq t_2.\]

			Next we will construct the set $\mathcal {K}$ by elements in $\Phi$.
			For each $\phi \in \Phi$, 
			we have the expression
			\[
			\begin{aligned}
			-XQ &= -M_T \mathcal {E}(\phi^\top \mysint B)_T 
			\\
			&= -M_t\mathcal {E}(\phi^\top \mysint B)_t - \int_t^T \mathcal {E}(\phi^\top \mysint B)_s\myd M_s - m_s \myd \mathcal {E}(\phi^\top \mysint B)_s - \myd \myangle{M, \mathcal {E}(\phi^\top \mysint B)}_s
			\\
			&= -M_t\mathcal {E}(\phi^\top \mysint B)_t - \int_t^T \mathcal {E}(\phi^\top \mysint B)_s\myd M_s - M_s \myd \mathcal {E}(\phi^\top \mysint B)_s -\sigma_s^X \phi_s \mathcal {E}(\phi \mysint B)_s \myd s
			\end{aligned} 	
			\]
			which implies
			\[
			\EEt t {-XQ} - \EEt t {-X} = \EEt t {\int_t^T \myangle{\sigma_s^X, -\phi_s \mathcal {E}(\phi \mysint B)_s}\myd s}.
			\]
			Taking the supremum, we have
			\[\begin{aligned}
			&\esssup_{Q\in \mathcal {Q}_t}\EEt t {-XQ} - \EEt t {-X} 
			\\
			=& \esssup _{\phi \in \Phi} \EEt t {\int_t^T \myangle{\sigma_s^X, -\phi_s \mathcal {E}(\phi^\top B)_s}\myd s}
			\\
			= &\EEt t {\int_t^T \sup_{\phi \in \Phi}\myangle{\sigma_s^X, -\phi_s \mathcal {E}(\phi^\top \mysint B)_s}\myd s}.
			\end{aligned} 
				\]
			The last equality holds true because the backward recursiveness of $(\mathcal {Q}_t)_{t\in [0,T]}$ implies the supremum can be taken from $T$ to $t$.

			Let $\mathcal {K}:= \{-\phi \mathcal {E}(\phi^\top \mysint B) \mid \phi \in \Phi\}$.
			By the correspondence, for each $t \in [0,T]$ we have 
			\[\begin{aligned}				
			\Dt t X = & \Ct t X - \EEt t X
			\\
			=&\esssup_{Q\in \mathcal {Q}_t}\EEt t {-XQ} - \EEt t {-X} 
			\\			
			=& \EEt t {\int_t^T \sup_{K \in \mathcal {K}}\myangle{\sigma_s^X, K_s}\myd s}.
			\end{aligned} 
			\] 
			Hence the time-consistent generalized deviation measure $\mathcal {D}$ can be represented by a robust volatility recorder.
		\end{myGoodProof*}	

		\begin{myGoodCorollary}
		{Time-consistent Volatility Recorder}
		{time-consistent-volatility-recorder}
		Given a robust volatility recorder $(\mathcal {V}_t)_t$ and the collection of weights $\mathcal {K}$, $\mathcal {V}$ is time-consistent if each element $K$ in $\mathcal {K}$ admits the form $K = -\phi \mathcal {E}(\phi^\top \mysint B)$  for some bounded $\phi$.
		\end{myGoodCorollary}
		
	\end{myblock}

	\begin{myblock}{Remark: Connections to $g$-deviation defined in \cite{pistorius2017dynamic}}
		The results in this section on time-consistent generalized deviation measures have some conceptual similarities to the work in \cite{pistorius2017dynamic}.
		However, there are a few differences behind the implementation.
		These differences are mainly manifested in two aspects: correspondence and time-consistency.
		It is the definition statement (D2'/D2't) and assumptions (A1)(A2) which cause the differences.
		Financial meanings of these statements and assumptions in this paper are clear and quite natural.
		Compared with the work in \cite{pistorius2017dynamic}, the correspondence and associated financial interpretations are proposed for coherent and generalized deviation measures in the conditional case.
		As for the time-consistency for generalized deviation measures, we start from a different point of view.
		We give the time-consistency in terms of the correspondence instead of generalizing conditional variance in \cite{pistorius2017dynamic}.
		It is the weak statement (D2't) and the associated time-consistency that allow us to give the mutual representation for time-consistent generalized deviation measures and $g$-deviations in Proposition~\ref{proposition:volatility-recorder-is-deviation}, Theorem~\ref{theorem:deviation-is-volatility-recorder} and Corollary~\ref{corollary:time-consistent-volatility-recorder}.
	\end{myblock}
\section{Risk Contribution}\label{Section:RiskContribution}
\subsection{What is Risk Contribution}
	\begin{myblock}{Risk Contribution: the Idea of `Hot Spot'}
		The idea of risk contribution comes from the sensitive analysis of the total risk. 
		In \cite{litterman1996hot} the asset that contributes to a large risk increment of the total risk is called the `hot spot'.
		The main role of risk contributions is to monitor how the total risk is composed and which part is more concentrated.
		Conceptually, the risk contributions of assets are considered as their share-scaled risk increments, i.e.,
		\[\text{Risk Contribution} = \text{Share} \times \text{Risk Increment}.\]
	\end{myblock}

	\begin{myblock}{Example: Static Standard Deviation Risk Contribution}
		The example following can be helpful to understand this concept.
		\begin{myGoodExample}
		{Risk Contribution - Static Standard Deviation}
		{single-period}
			Suppose there are $d$ assets with their $\mathbb R^d$-valued random returns $r$ and the positive definite covariance matrix is expressed as $\Lambda \in \mathbb R^{d\times d}$.  
			We use $w\in \mathbb R^d$ to denote the weights on these assets, which describes how the total capital is allocated.
			We set the standard deviation as the risk measure to quantify the risk of this portfolio.
			The total risk under the capital allocation $w$ is denoted as
			\[f(w) := \sqrt{w^\top \Lambda w} = \sum_{j = 1}^d w_j \dfrac{\partial f}{\partial w_j}.\]
			The $i$-th asset's infinitesimal increment is the partial derivetive $\dfrac{\partial f}{\partial w_i} $ and is called the marginal risk contribution of $i$-th asset.
			The risk contribution is then defined as $w_i \dfrac{\partial f}{\partial w_i}$.
			The positive homogeneity of $f$ implies the aggregation property
			\[f(w) = \sqrt{w^\top \Lambda w} = \dfrac{1}{\sqrt{w^\top \Lambda w}}\sum_{j = 1}^d w_j \left(\Lambda w\right)_j =\sum_{j = 1}^dw_j\dfrac{\partial f}{\partial w_j} \]
			which means the total risk is the sum of individual risk contributions.
		\end{myGoodExample}
	\end{myblock} 
\subsection{Risk Contribution for Static/Conditional Measures}
	\begin{myblock}{Risk Contribution and Outcome of the Measurements in the Continuous-time Case}
		The risk contribution should be distributed in the interval $[0,T]$ if the investments are implemented continuously over this time horizon, even with a static risk measure for terminal wealth.
		Hence we only need to investigate the risk contributions for conditional risk measures for that static measures are special cases.

		Now consider a predictable set $E\in \Sigma_p$, its associated indicator process $\II_E$ can be recognized as the policy that holds a unit share of assets on $E$.
		Based on the policy $u$, the risk increment brought by the extra share $\II_E$ is
		\[\rho(X_T^{u+ \II_E}) - \rho(X_T^u)\]
		for a conceptual risk measure $\rho$.
		The infinitesimal version of this increment can be recognized as the ``hot spot'' of the policy $u$ on the set $E$, namely
		\[\lim_{\theta \to 0}\dfrac{\rho(X_T^{u+\theta \II_E}) - \rho(X_T^u)}{\theta}.\]
		We can get an understanding of continuous-time risk contribution from the expression above informally.
		Here we state several lemmata before giving the formal definition of continuous-time risk contribution.
	\end{myblock}

	\begin{myblock}{Exposed Face}
		Given the risk envelope $\mathcal {Q}_t$ with its associated conditional coherent risk measure $\mathcal {C}_t$ and generalized deviation measure $\mathcal {D}_t$, the \emph{exposed face} of a financial position $X$ is defined as
		\[\mathcal {R}_t (X ):= \left\{ \widehat Q \in \mathcal {Q}_t \Big | \Ct{t}{X} = \EEt{t}{-XQ}\right\}.\]
		The elements in the exposed face $\mathcal {R}_t(X)$ are usually considered as the worst case in $\mathcal {Q}_t$ when we take the robust expectation of $-X$.
		They also play an instrumental role in computing the subdifferential of risk measures.
		Inspired by \cite{cherny2009capital}, we give a refined version for conditional risk measures in the following lemma.
	\end{myblock}
	
	\begin{myblock}{Subdifferential}			
		\begin{myGoodLemma}
		{Subdifferential}
		{subdifferential}
			For $X, Y \in L^2_T$, we have
			\[\lim_{\theta \to 0} \sup \dfrac{\Ct{t}{X+\theta Y} - \Ct{t}{X} }{\theta} = \esssup _{\widehat Q \in \mathcal {R}_t(X)}\EEt{t}{-\widehat QY},\]
			\[\lim_{\theta \to 0} \inf \dfrac{\Ct{t}{X+\theta Y} - \Ct{t}{X} }{\theta} = \essinf _{\widehat Q \in \mathcal {R}_t(X)}\EEt{t}{-\widehat QY},\]
			\[\lim_{\theta \to 0} \sup \dfrac{\Dt{t}{X+\theta Y} - \Dt{t}{X} }{\theta} = \esssup _{\widehat Q \in \mathcal {R}_t(X)}\EEt{t}{(1-\widehat Q)Y},\]
			\[\lim_{\theta \to 0} \inf \dfrac{\Dt{t}{X+\theta Y} - \Dt{t}{X} }{\theta} = \essinf _{\widehat Q \in \mathcal {R}_t(X)}\EEt{t}{(1-\widehat Q)Y}.\]
		\end{myGoodLemma}
	\end{myblock}
	\begin{myGoodProof*}{}{}
		For the correspondence to generalized deviation measures, we only need to show the results of conditional coherent risk measures.
		Now consider the functional $\rho: L_T^2 \to \mathbb {R}$ defined as
		\[\rho (X):= \mathbb {E}\left[\Ct{t}{X}\right].\]
		This functional is a coherent risk measure and has the dual representation
		\[\rho (X) = \sup_{Q\in \mathcal {Q}}\mathbb {E}[-XQ].\]
		The associated exposed face is denoted as $\mathcal {R}(X)$.

		First, we introduce the set $G$ by 
		\[G = \closed \left\{(\mathbb {E}[-XQ], \mathbb {E}[-YQ]) \Big| Q\in \mathcal {Q}\right\}.\]
		This set $G$ is convex and closed in $\mathbb {R}^2$.
		We can also see that
		\begin{equation}
		\begin{aligned}
		\rho ({\alpha X + \beta Y}) &= \sup _{Q\in \mathcal {Q}}\mathbb {E}[-Q(\alpha X + \beta Y)]
		\\
		&=\sup _{Q\in \mathcal {Q}}\myangle{(\alpha, \beta), (\mathbb {E}[-QX], \mathbb {E}[-QY])}
		\\
		&= \max _{(x,y)\in G} \myangle{(\alpha, \beta),(x,y)}.
		\end{aligned}
		\end{equation}
		Then we denote $a = \min \{x| \exists x \text{ suth that }(x,y)\in G\}$, $\overline{b} = \max \{y | (a,y)\in G\}$ and $\underline{b} = \min \{y | (a,y)\in G\}$ for convenience. 

		Now we let $\alpha = 1, \beta = \theta$.
		For $\theta >0$, the maximum $\max _{(x,y)\in G} \myangle{(1 , \theta),(x,y)}$ can be attained at a point $(a_\theta, b_\theta)$ such that 
		\begin{itemize}
				\item $a_\theta \leq a$ and $a_\theta \to a$ as $\theta \downarrow 0$;
				\item $b_\theta \geq \overline{b}$ and $b_\theta \to \overline{b}$ as $\theta \downarrow 0$;
				\item $a_\theta + \theta b_\theta \geq a+ \theta \overline{b} \geq a + \theta \underline{b} $.
		\end{itemize}
		For $\theta < 0$, we can get the following properties respectively:
		\begin{itemize}
				\item $a_\theta \leq a$ and $a_\theta \to a$ as $\theta \uparrow 0$;
				\item $b_\theta \leq \overline{b}$ and $b_\theta \to \underline{b}$ as $\theta \uparrow 0$;
				\item $a_\theta + \theta b_\theta \geq a+ \theta \underline{b} \geq a + \theta \overline{b} $.
		\end{itemize}

		As a result, 
		\begin{equation}
			\label{eq:sup-variation}
			\begin{aligned}
			\lim_{\theta \to 0} \sup \dfrac{\rho ({X + \theta Y}) - \rho (X)}{\theta} &=  \lim_{\theta \to 0} \sup \left[\sup _{(x,y)\in G}(x+ \theta y) -  \sup _{(x,y)\in G}x\right]
			\\
			&=\lim _{\theta \to 0}\sup \dfrac{1}{\theta}\left[ a_\theta + \theta b_\theta - a \right]
			\\
			&= \overline{b} = \sup _{Q \in \mathcal {R}(X)}\mathbb {E}[-QY]
			\end{aligned}
		\end{equation}
		for 
		\[b_\theta \geq \dfrac{a_\theta - a}{\theta} + b_\theta \geq \underline{b}.\]

		Noticing that the elements in $\mathcal {R}_t(X)$ are consistent on $\mathcal {F}_t$, we have 
		\begin{equation}
		\label{eq:esssup-variation-A}
		\begin{aligned}
			&\lim_{\theta \to 0} \sup \dfrac{\rho ({\II_A(X + \theta Y)}) - \rho (\II_AX)}{\theta}	 
			\\
			=& \sup_{Q\in \mathcal {R}_t(X\II_A)}\mathbb {E}[-YQ\II_A]
			\\
			=& \sup_{Q\in \mathcal {R}_t(X)}\mathbb {E}[-YQ\II_A]
			\\
			=& \mathbb {E}\left[\II_A\esssup_{ Q \in \mathcal {R}_t(X)}\EEt{t}{-YQ}\right]
		\end{aligned} 
		\end{equation}
		where the first equation is implied by equation~(\ref{eq:sup-variation}).

		Using the techniques in Theorem~\ref{theorem:conditional-correspondence}, 
		we also have
		\begin{equation}
			\label{eq:esssup-variation-B}
		\begin{aligned}
			&\lim_{\theta \to 0} \sup \dfrac{\rho ({\II_A(X + \theta Y)}) - \rho (\II_AX)}{\theta}
			\\
			=& \lim_{\theta \to 0}\sup \dfrac{1}{\theta} \mathbb {E}\left[\II_A \esssup_{Q\in \mathcal {Q}_t}\EEt{t}{-(X+\theta Y)Q} - \II_A \esssup_{Q\in \mathcal {Q}_t}\EEt{t}{-XQ}\right].
		\end{aligned} 
		\end{equation}

		Comparing equation~(\ref{eq:esssup-variation-A}) with equation~(\ref{eq:esssup-variation-B}) for arbitrary $A\in \mathcal {F}_t$, we can get the first statement of this lemma.
		The second statement lim-inf result can be given in the same manner.
	\end{myGoodProof*}

	\begin{myblock}{{\Doleans} Process}
		\begin{myGoodLemma}
		{{\Doleans} Measure}
		{doleans-process}
			With the fixed time $t^\star$ and an arbitrary element $\widehat Q \in \mathcal {R}_{t^\star}(X_T^u)$, the set function $E\mapsto \mathbb {E}[-X_T^{\II_E}\widehat Q]$ is a signed measure on the predictable $\sigma$-algebra $\Sigma_p$ of $\Omega \times [0,T]$.
			This measure denoted as $\mu$ can be expressed uniquely by a process $\ell$ by 
			\begin{equation}
				\mu (E) = \mathbb {E}\int_0^T \II_E\ell_t \myd t.
			\end{equation}
		\end{myGoodLemma}
	\end{myblock}
	Suppose $Q$ is the {\RadonNikodym} derivative of $\mathbb {Q}$ with respect to $\mathbb {P}$ with the expression $Q = \mathcal {E}(-\theta \mysint W)_T$ where $\mathcal {E}$ is the {\Doleans}-Dade exponential.
	We say that $\theta$ is the price of risk under $\mathbb {Q}$.
	We can rewrite the $\mathbb {Q}$-expected loss of $X_T^u$
	\[
	\begin{aligned}
		\mathbb {E}[-X_T^u Q] &= \mathbb {E}\left[-X_T^u \mathcal {E}(-\theta \mysint W)_T\right]
		\\
		&=-\mathbb {E}\left[\int_0^T u_t^\top (b_t - \sigma_t \theta _t^\top )\mathcal {E}(-\theta \mysint W)_t\myd t\right]
		\\
		&= \mathbb {E^Q}\left[\int_0^T-u_t^\top (b_t - \sigma_t \theta_t^\top) \myd t\right].
	\end{aligned}  
	\]
	Under $\mathbb {Q}$ , the term $-u_t^\top (b_t - \sigma_t \theta_t^\top)$ can be considered as the instantaneous loss of $X^u$ and the term $-b_t + \sigma_t \theta_t^\top$ can also be recognized as the risk-adjusted loss of assets.
	As a special case, when we take the risk-neural probability, the risk-adjusted loss is zero.

	Compared with the expression above, the process $\ell$ defined in Lemma~\ref{lemma:doleans-process} is the risk-adjusted loss of assets under $\widehat {\mathbb {Q}}$ introduced by $\myd \widehat{\mathbb {Q}} = \widehat Q \myd \mathbb {P}$.
	Notice that the process $\ell$ is introduced by the elements in $\mathcal {R}_t(X_T^u)$ and it can be interpreted as the worst-case instantaneous loss.

	Alternatively, if we replace $X$ by $X_T^u$ and $Y$ by $X_T^{\II_E}$ in Lemma~\ref{lemma:subdifferential}, $\mu (E)$ lies between sup/inf marginal risk increment under the policy $u$.
	Hence, the process $\ell$ also refers to marginal risk increments.
	The following proposition shows that share-scaled marginal risk increment admits an aggregation property.

	\begin{myblock}{Aggregation}
		\begin{myGoodProposition}
		{Aggregation}
		{aggregation}
			With the same $t^\star$ and $\widehat Q \in \mathcal {R}_{t^\star}(X_T^u)$ in Lemma~\ref{lemma:doleans-process}, the process $\ell$ in Lemma~\ref{lemma:doleans-process} satisfies the following properties:
			\begin{enumerate}
				\item $\mathbb {E}\int_0^T v_t \ell_t \myd t = \mathbb {E}[-X_T^v\widehat Q]$;
				\item $\EEt{t^\star}{\int_{t^\star}^T v_t \ell_t \myd t} =  \Ct{t^\star}{X_T^v-X_{t^\star}^v}$.
			\end{enumerate}
			As a consequence, the conditional measures at time $t^\star$ admit the aggregation property
			\[\Ct{t^\star}{X_T^u} = \EEt{t^\star}{\int_{t^\star}^T v_t \ell_t \myd t} \Big |_{v = u} - X^u_{t^\star},\]
			\[\Dt{t^\star}{X_T^u} = \EEt{t^\star}{\int_{t^\star}^T v_t \ell_t \myd t} \Big |_{v = u} + \EEt{t^\star}{X^u_{T} - X^u_{t^\star}} = \EEt{t^\star}{\int_{t^\star}^T v_t (\ell_t+b_t) \myd t} \Big |_{v = u} .\]
		\end{myGoodProposition}
	\end{myblock}

	\begin{myblock}{Define the Risk Contribution}
		\begin{myGoodDefinition}
		{Risk Contribution for Conditional Measures}
		{conditional-contribution}
			With the same $t^\star$ and $\widehat Q \in \mathcal {R}_{t^\star}(X_T^u)$ in Lemma~\ref{lemma:doleans-process}, the \emph{marginal risk contribution} of the policy $u$ under $\Ct{t^\star}{\cdot}$ is defined as the process $\ell$ 
			\[m_t^{\mathcal {C}_{t^\star}} = \ell_t \text{ for }t\in [t^\star,T]\]
			and that under $\Dt{t^\star}{\cdot}$ is defined as $\ell+b$
			\[m_t^{\mathcal {D}_{t^\star}} = b_t +\ell_t \text{ for }t\in [t^\star,T].\]
			The associated risk contribution $c^{\mathcal {C}_{t^\star}}$ is 
			defined by 
			\[c^{\mathcal {C}_{t^\star}}_t = u_t \odot m^{\mathcal {C}_{t^\star}}_t \text{ for }t\in [t^\star,T]\]
			where $\odot$ is the element-wise multiplication for vectors.
		\end{myGoodDefinition}
		Hence by Proposition~\ref{proposition:aggregation}, the conditional coherent risk measure can be expressed as
		\begin{equation}
		\label{eq:coherent-aggregation}
		\Ct{t^\star}{X_T^u} = \EEt{t^\star}{\int_{t^\star}^Tu_tm^{\mathcal {C}_{t^\star}}_t \myd t} -X_{t^\star}^u
		\end{equation}
		and 
		the conditional generalized deviation measure can be expressed as
		\begin{equation}
		\label{eq:deviation-aggregation}
		\Dt{t^\star}{X_T^u} = \EEt{t^\star}{\int_{t^\star}^Tu_tm^{\mathcal {D}_{t^\star}}_t \myd t}.
		\end{equation}
		The equations (\ref{eq:coherent-aggregation}) and (\ref{eq:deviation-aggregation}) can be recognized as the decomposition of the total risk 
		\begin{equation}
		\label{eq:idea-aggregation}
		\underbrace{\rho_{t^\star}(X_T^u - X_{t^\star}^u)}_{\text{total risk}}   
		= 
		\EEt{t^\star}{\int_{t^\star}^T  
		\sum_{i = 1}^d
		\overbrace{
		\underbrace{u_t^i}_{\text{shares}} 
		\underbrace{m_t^i}_{\text{marginal}} 
		}
		^{\text{risk contribution}}
		\myd t}.
		\end{equation}
		which can also be called the aggregation of risk contribution.
	\end{myblock}
	
	\begin{myblock}{Explanation of Conditional Risk Contributions}
		These two types of measures, i.e. coherent risk measures and generalized deviation measures, behave differently in terms of risk contribution.
		For coherent risk measures, the marginal risk contribution can be interpreted as the worst-case instantaneous loss of assets and the risk contribution is the shares-scaled worst-case loss induced.
		As for conditional generalized deviation risk measures, the marginal risk contribution is recognized as the difference of the worst-case instantaneous loss $\ell (\omega, t)$ of assets and the referenced instataneous loss $-b(\omega,t)$.
		Respectively, the risk contribution of generalized deviation measures is the shares-scaled difference and can also be considered as the instataneous `amplitude'.	  
		Interestingly, the (marginal) risk contribution under $\mathcal {D}_{t^\star}$ does not charge until time $t^\star$ --- what has happened till $t^\star$ is constancy in the view of $\mathcal {D}_{t^\star}$ and hence does not contribute to the total risk.
		If the outcomes of coherent/deviation risk measures are recognized in a macro view, their associated risk contributions are the micro components of the total risk.
		In a microscope view, their risk contributions also have a correspondence with each other.			
	\end{myblock}

\section{Risk Contributions for Time-consistent Measures}\label{Section: Time-Consistent Risk Contribution}
\subsection{Time-consistency of Risk Contribution}
	\begin{myblock}{The Idea of Time-consistent Risk Contributions}
		When given a $\mathcal {F}_{t^\star}$-conditional risk measure, we can check the ``hot spot'' of the total risk through the risk contribution defined in Section~\ref{Section:RiskContribution}.
		However, we may be in trouble when given risk measures conditional at different time points, for example $\mathcal {C}_{t_1}$ and $\mathcal {C}_{t_2}$.
		In the microscope view, the ``hot spots'' for risk measures conditional at different time points may not be consistent.	
		That leads now to an interesting question: is risk contribution time-consistent for time-comsistent risk measures?
		This statement is equivalent to saying that the marginal risk contribution $m^{\Ct{t_1}{X_T^u}}$ is identical to $m^{\Ct{t_2}{X_T^u}}$ for each $t_1,t_2$ (or respectively for $(\mathcal {D}_t)_t$) when $\mathcal {C}$ or $\mathcal {D}$ is time-consistent.

		Now suppose the time-consistent risk measures $(\mathcal {C}_t)_t$ and $(\mathcal {D}_t)_t$ admit the correspondence $\Dt{t}{\cdot} = \Ct{t}{\cdot} + \EEt{t}{\cdot}$ for each $t\in [0,T]$ and the wealth process of the investment $X^u$ with the policy $u$ is 
		\begin{equation}
		\label{eq:investment-wealth}
		\myd X^u_t = u_t ^\top b_t \myd t + u_t^\top \sigma_t \myd B_t, X_0 = x.
		\end{equation}
		The marginal risk contribution of $\mathcal {C}_{t^\star}$ is denoted as $m^{\mathcal {C}_{t^\star}}$ (and $m^{\mathcal {D}_{t^\star}}$ for the generalized deviation measure respectively) for each $t^\star$.

		\begin{myGoodTheorem}
		{Time-consistent Risk Contribution}
		{time-consistent-risk-contribution}
		The statements are true for time-consistent risk measures $\mathcal {C}$ and $\mathcal {D}$.
			\begin{enumerate}
				\item The marginal risk contribution of policy $u$ is time-consistent, i.e., 
				\[m_t^{\mathcal {C}_{t_1}} = m_t^{\mathcal {C}_{t_2}}, m_t^{\mathcal {D}_{t_1}} = m_t^{\mathcal {D}_{t_2}}\]
				for each $t\geq t_1\vee t_2$ with arbitrary $t_1$ and $t_2$.
				Their associated risk contributions are also time-consistent.
				\item The marginal risk contributions of these two measures have the correspondence
				\[m^{\mathcal {D}} = m^{\mathcal {C}} + b.\]
				\item Suppose that the time-consistent generalized deviation measure is represented by $\mathcal {K}$
				\[\Dt{t^\star}{X_T^u} = \EEt{t^\star}{\int_{t^\star}^T\sup_{K\in \mathcal {K}}\myangle{K_t, u_t^\top \sigma_t}\myd t}.\]
				The marginal risk contribution $m^{\mathcal {D}}$ can be identified by 
				\[m^{\mathcal {D}}(\omega,t) =  \sigma(\omega,t)\widehat{K}(\omega,t) \]
				where $\widehat K_t$ satisfies $  \sup_{K\in \mathcal {K}}\myangle{K_t, u_t^\top \sigma_t} = u_t^\top \sigma_t\widehat K_t$.
			\end{enumerate}
		\end{myGoodTheorem}
		\begin{myGoodProof*}{}{}
			With fixed time $t^\star$, by taking the variation of $\Dt{t^\star}{X_T^u}$ on the policy $u$, we can get the marginal risk contribution 
			\[m^{\mathcal {D}_{t^\star}}_t =\sigma_t \widehat{K}_t\]
			for each $t\geq t^\star$
			where $\widehat K_t$ is one of the subgradients of the function $u_t \mapsto \sup_{K\in \mathcal {K}}\myangle{K_t, u_t^\top \sigma_t}$ for every $(\omega, t)$.
			Any subgradient $\widehat K_t$ satisfies 
			\[\sup_{K\in \mathcal {K}}\myangle{K_t, u_t^\top \sigma_t} = u_t^\top \sigma_t\widehat K_t.\]
			For arbitrary fixed $t_1$ and $t_2$, the marginal contribution for $\mathcal {D}_{t_1}$ and $\mathcal {D}_{t_2}$ satisfy the equation
			\[m_t^{\mathcal {D}_{t_1}} = \sigma_t \widehat{K}_t  =m_t^{\mathcal {D}_{t_2}}\]
			for each $t\geq t_1\vee t_2$.
			Hence the marginal risk contribution for the time-consistent generalized deviation risk measure $\mathcal {D}$ can be denoted as $m^{\mathcal {D}}$ in brief.
			We get the time-consistency of risk contribution for robust volatility recorders in the third statement and the second equality in the first assertion.
			
			The correspondence of the conditional risk contributions in Definition~\ref{definition:conditional-contribution} shows that
			\[m_t^{\mathcal {D}_{t^\star}} = m_t^{\mathcal {C}_{t^\star}} + b_t\]
			for each $t^\star$.
			With the time-consistency for $m^{\mathcal {D}}$, the equation above implies the first equality in the first statement and the second statement.
		\end{myGoodProof*}
	\end{myblock}
\subsection{Time-consistent Risk Contribution and \texorpdfstring{$g$}{g}-expectation}\label{Subsection: Z-is-risk-contribution}
	\begin{myblock}{$Z$ as the Risk Contribution}
		Recall that time-consistent coherent risk measures are always connected to the nonlinear expectation $g$-expectation, which is the component $(Y_t)_t$ of the solution $(Y,Z)$ to the BSDE
		\begin{equation}
			\label{eq:BSDE-risk-contribution}
			\begin{cases}
					-\myd Y_t = g_t(Z_t) \myd t - Z_t \myd B_t
					\\
					Y_T = -X_T^u.
			\end{cases}
		\end{equation}
		In \cite{gianin2006risk}, the first component $Y$ of the solution pair $(Y,Z)$ is recognized as the outcome of its associated coherent risk measure $(\mathcal {C}_t)_t$.
		However, there is almost no discussion on the financial interpretation of the second component $Z$.
		In fact, the second solution component $Z$ is closely related to the risk contribution, when taking the risk measure of the terminal wealth of investments.
		
		Suppose the wealth of the investment $X^u$ is still expressed by the dynamic (\ref{eq:investment-wealth}).
		Solve BSDE~(\ref{eq:BSDE-risk-contribution}) above and we have
		\[
			\EEt{t^\star}{Y_{t^\star} - Y_T} = \EEt{t^\star}{\int_{t^\star}^T g_t(Z_t)\myd t}  = \EEt{t^\star}{\int_{t^\star}^T \sup_{\phi \in \Phi }\myangle{\phi_t, Z_t}\myd t} = \EEt{t^\star}{\int_{t^\star}^T \myangle{\widehat\phi_t, Z_t}\myd t}.
		\]
		Now consider the deviation risk measure $\mathcal {D}$ corresponded to $\mathcal {C}$.
		Comparing the expression with the construction of $\mathcal {K}$ in Theorem~\ref{theorem:deviation-is-volatility-recorder}, we can get 
		\[\begin{aligned}
			\Dt{t^\star}{X_T^u} = &\EEt{t}{\int_{t^\star}^T\sup_{K\in \mathcal {K}}\myangle{K_t, \sigma^{X^u}_t}\myd t} 
			\\
			=& \EEt{t^\star}{\int_{t^\star}^T\myangle{\widehat K_t, \sigma^{X^u}_t}\myd t} 
			\\
			=& \EEt{t^\star}{\int_{t^\star}^T\myangle{-\widehat \phi \mathcal {E}(\widehat \phi^\top\mysint  B)_t, \sigma^{X^u}_t}\myd t}
			\\
			=& \EEt{t^\star}{\int_{t^\star}^T\myangle{\widehat \phi , -\sigma^{X^u}_t\mathcal {E}(\widehat \phi^\top\mysint  B)_t}\myd t}
			\end{aligned} 
		\]
		which implies
		\[Z_t = -\sigma_t^{X^u}\mathcal {E}(\widehat \phi \mysint B)_t = -u^\top_t\sigma_t\mathcal {E}(\widehat \phi \mysint B)_t.\]

		On the one hand, $Z$ can be interpreted as the volatility under the worst-case probability $ \mathbb {\widehat Q}$ introduced by $\mathcal {E}(\widehat \phi ^\top \mysint B)_T$ from the last equality.
		On the other hand, noticing the relation between $\mathcal {E}(\widehat \phi ^\top \mysint B)$ and $\widehat K$, we will give another interpretation of $Z$ in terms of  the risk contribution of $\mathcal {D}$.
		Now consider the marginal risk contribution $m_t^\mathcal {D} = \sigma_t\widehat K_t$ and the risk contribution
		\[c^\mathcal {D} = u_t\odot m_t^\mathcal {D} =u_t\odot \left(\sigma_t\widehat K_t\right)\]
		at time $t$.
		The summation risk contribution of all assets can be calculated as
		\[\sum_{i = 1}^d c_t^{\mathcal {D},i} = \myangle{u_t, \sigma_t\widehat K_t} = Z_t \widehat \phi_t = Z_t \dfrac{\partial g_t(z)}{\partial z}\Big |_{z = Z_t}.\]		
		If we take the Brownian motion as the risk source factor and the volatility $\sigma_t$ as the factor loading of assets, $u^\top\sigma$ can be considered as a recombination of factor loadings according to the policy $u$.
		The generalized deviation risk measure $\mathcal {D}$ acts as the robust volatility recorder for this recombined factor loading with the candidate weights $\mathcal {K}$.
		When we shed light on the risk contribution $c^\mathcal {D}$, the solution $Z_t$ with the worst-case kernal $\phi$ which can also expressed by $\dfrac{\partial g_t(z)}{\partial z}$, can give the summation of assets' risk contribution.
		
		\end{myblock}

\subsection{Example}
	\begin{myblock}{Example: Chen-Epstein $\kappa$-ignorance}
		Now we give an example to illustrate the correspondence and risk contribution for time-consistent coherent and generalized deviation risk measures.
		We start from constructing a robust volatility recorder and finally derive the $\kappa$-ignorance established in \cite{chen2002ambiguity} in the view of correspondence.
		$\kappa$-ignorance is a manner of robust pricing when investors are facing the ambiguity of probabilities under which the price of risk is uncertain and $[-\kappa, \kappa]$-valued.

		Suppose the wealth process of the investment $X^u$ with the policy $u$ satisfies the dynamic~(\ref{eq:investment-wealth}).
		Under the reference probability $\mathbb {P}$, we can define a robust volatility recorder for the position $X_T^u$ by
		\[\Dt{t^\star}{X_T^u} :=\EEt{t^\star}{\int_{t^\star}^T \sum_{j = 1}^m \sup_{K \in \mathcal {K}}\myangle{K_s, u_s^\top \sigma_s}\myd s} \text{ for $t^\star \in [0,T]$}\]
		where $\mathcal {K}$ is given by $\mathcal {K}=\{\phi \mathcal {E}(-\phi^\top \mysint B)\mid \phi \in [-\kappa, \kappa]\}$.
		For $\phi \in [-\kappa, \kappa]$, the volatility $\sigma^{X^u}$ in the view of the probability introduced by $\mathcal {E}(-\phi^\top \mysint B)_T$ is recorded as $ \myangle{\phi_t , u_t^\top \sigma_t}\mathcal {E}(-\phi^\top \mysint B)_t$.
		The kernals of these stochastic exponentials, which can also be understood as the price of risk, is bounded by a constant $\kappa$.
		And $\mathcal {D}$ is the robust version of these candidate recorders.

		By the correspondence in Theorem~\ref{theorem:conditional-correspondence}, we can derive the associated coherent risk measure by
		\[\Ct{t^\star}{X_T^u} = \EEt{t^\star}{-X_T^u} + \Dt{t^\star}{X_T^u}.\]
		According to Subsection~\ref{Subsection: Z-is-risk-contribution},  $\mathcal {C}$ is also connected to the solution to the BSDE
		\[
		\begin{cases}
		-\myd Y_t = \myangle{\kappa 1_d, \myvert{Z_t}} \myd t - Z_t \myd B_t
		\\
		Y_T = -X_T^u.
		\end{cases} 	
		\]
		$\mathcal {C}_{t^\star}$ can also be represented by
		\[\Ct{t^\star}{X_T^u} = \esssup_{Q\in \mathcal {Q}}\EEt{t^\star}{-X_T^uQ}\]
		where $\mathcal {Q} = \{\mathcal {E}(\phi \mysint B)_T\mid \phi \in [-\kappa, \kappa]\}$.
		This is a manner of pricing the loss which is called $\kappa$-ignorance in the view of \cite{chen2002ambiguity}.
		Here this $\kappa$-ignorance pricing manner can be derived from the robust volatility recorder.
		The constant $\kappa$ acts as volatility weight in $\mathcal {D}$ while it measures the ambiguity when pricing financial positions.
		When $\kappa = 0$, the recorder will not record any volatility from the dynamic of the investment $X^u$, i.e. $\Dt{t^\star}{X_T^u}\equiv 0$.
		That is to say the difference between coherent measure $\Ct{t^\star}{X_T^u}$ and reference expected $\EEt{t^\star}{-X_T^u}$ loss is zero.
		This is consistent with the view about pricing under ambiguity in \cite{chen2002ambiguity}.		
		Moreover, the sum of $\mathcal {D}$ risk contributions under $\mathbb {P}$ of all assets at time $t$ is given by
		\[u_t^\top m_t^{\mathcal {D}} = \kappa \myangle{Z_t, \sign (Z_t)} = \kappa \myangle{1_d, |u_t^\top \sigma_t|}\mathcal {E}(\widehat \phi \mysint B)_t.\]
		In other words, under the worst-case $\mathbb {\widehat Q}$, the $\mathcal {D}$ risk contribution is recorded as the accumulation of $\kappa \myangle{1_d, |u_t^\top \sigma_t|}$.
	\end{myblock}

\section{Conclusion}\label{Section: Conclusion}
	\begin{myblock}
	{Conclusion}
		We summarize the contributions of this paper as follows
		\begin{itemize}
			\item Give the definition for conditional generalized deviation measures.
			Based on this axiomatic description, we also show the correspondence between conditional generalized deviation measures and coherent risk measures which is the extension of the static case in \cite{rockafellar_generalized_2006} to the continuous-time case.
			\item According to the time-consistency of coherent risk measures, we give a description of the time-consistency of generalized deviation measures via the tool of volatility recorders and $g$-expectation.
			\item The risk contribution is characterized for conditional risk measures. The expression (\ref{eq:idea-aggregation}) gives core idea of conditional risk contributions.
			As a special case, the risk contribution of time-consistent measures is proved to be time-consistent.
			\item We also link the solution $(Y,Z)$ to the risk characterizations.
			$Y$ is recognized as the outcome of the time-consistent coherent risk measures.
			$Z$ can be interpreted as the risk contribution of the generalized deviation measure introduced from the coherent measure.	  
		\end{itemize}
	\end{myblock}

\bibliographystyle{apalike}
\bibliography{MyBib}
\end{document}